\documentclass[10pt,a4paper,twoside]{revciuni}
\renewcommand{\figurename}{Fig.}
\usepackage{anysize} 
\marginsize{1.5cm}{1.5cm}{0.5cm}{1cm}
\usepackage{latexsym}
\usepackage[utf8]{inputenc}
\usepackage{amsmath,amsthm,amsfonts,amssymb,amscd,euscript,amsfonts}
\usepackage{capt-of}
\usepackage{multicol}
\usepackage{graphicx}
\usepackage{dcolumn}
\usepackage{bm}%

\usepackage[calcwidth]{titlesec}
\newpagestyle{estiloDa}[\small]{\headrule
\sethead[\usepage][Pereyra et al.][]{}{OAUNI: first light}{\thepage}
\footrule\setfoot[Facultad de Ciencias -- UNI][REVCIUNI XX (X) (2015) XX--XX][]{}{REVCIUNI XX (X) (2015) XX--XX}{Facultad de Ciencias -- UNI}}
\newpagestyle{primeraDa}{\headrule
\sethead[][][]{}{\large{Revista de la Facultad de Ciencias de la UNI, REVCIUNI XX (X) (2015) XX--XX}}{}
\footrule\setfoot[][][]{}{Facultad de Ciencias -- Universidad Nacional de Ingeniería}{}}

\begin{document}

\title{UNI Astronomical Observatory - OAUNI: \\ First light}

\author{Antonio Pereyra\supit{\rm{1}}, Julio Tello\supit{\rm{2}}, Erick Meza\supit{\rm{2,3}}, William Cori\supit{\rm{2}}, \\ José Ricra\supit{\rm{2}}, Maria Isela Zevallos\supit{\rm{2}}\\
\supit{\rm{1}}\it{Instituto Geofísico del Perú, Área Astronomía\\
\supit{\rm{2}}\it{Grupo Astronomía, Facultad de Ciencias, Universidad Nacional de Ingeniería}\\
\supit{\rm{3}}\it{l'Observatoire de Paris, France}\\
\it{apereyra@igp.gob.pe}\\
}}

\date{Submitted 2015/06/11; accepted 2015/06/17}      
\maketitle
\thispagestyle{primeraDa}

\hfill

\begin{abstract}

We show the actual status of the project to implement the Astronomical Observatory of the National University of Engineering (OAUNI), including its first light. The OAUNI was installed with success at the site of the Huancayo Observatory on the peruvian central Andes. At this time, we are finishing the commissioning phase which includes the testing of all the instruments: optical tube, robotic mount, CCD camera, filter wheel, remote access system, etc. The first light gathered from a stellar field was very promissory. The next step will be to start the scientific programs and to bring support to the undergraduate courses in observational astronomy at the Faculty of Sciences of UNI. \\


\end{abstract}

\hfill
\hfill

\begin{multicols}{2} 

\section{Introduction}\label{intro}

The astronomical observatory project for the National University of Engineering (OAUNI,~\cite{mez09,per12}) began in 2009. The initiative and strong support of the Astronomy Group (GA) at the Faculty of Sciences (FC$-$UNI) was fundamental for this purpose. Quickly the project gains additional support of the peruvian professional astronomers working in other countries, being the majority of them ex-members of GA.

The main aim of this project is to implement a facility to perform astronomical observations for the FC-UNI. This will let that scientific projects and graduate theses can be developed in observational astronomy. Following this line, OAUNI will give support to the courses in astronomy actually being imparted at FC-UNI. Between the scientific programs priority will be given to site testing ~\cite{per03,bae03,dal04,mez13,mez15}, stellar photometry including variability of middle and large term, stellar spectroscopy~\cite{dal07} and astronomical polarimetry. Astronomical transient events as supernovas, novas and stellar occultations~\cite{mez15} also will be considered. In addition, astronomy outreach also will be benefited with OAUNI.

At the last years the project has been acquiring the basic equipment and instruments to complete the observatory. Since its beginning, the project had financial support of several sources. In 2011, the UNI Rectorate gave support for the acquisition of the optical tube for the scope. The importation of this equipment was completed at the end of 2011. New financial sources were explored and in 2013 the project obtained a research grant from \textit{The World Academy of Sciences} (TWAS). This budget let us to acquire the robotic mount and a new CCD detector in 2014. Additional custom expenses were assumed by the UNI Rectorate and FC$-$UNI. Finally, a grant for a master student with a thesis associated to OAUNI project also was approved by TWAS.

Up to now, the main instruments of OAUNI are a 0.5m optical tube of extreme quality, a precision robotic mount, a couple of CCD detectors, a spectrograph, and an astronomical filters set with its own automatic filter wheel. On the first semester of 2014, we completed the pre-comiisioning phase that included the engage of the optical tube over the robotic mount. These indoor tests were  inside the FC$-$UNI building. 

\begin{figure*}
\begin{picture}(0,180)
    \centerline{\includegraphics[width=14cm]{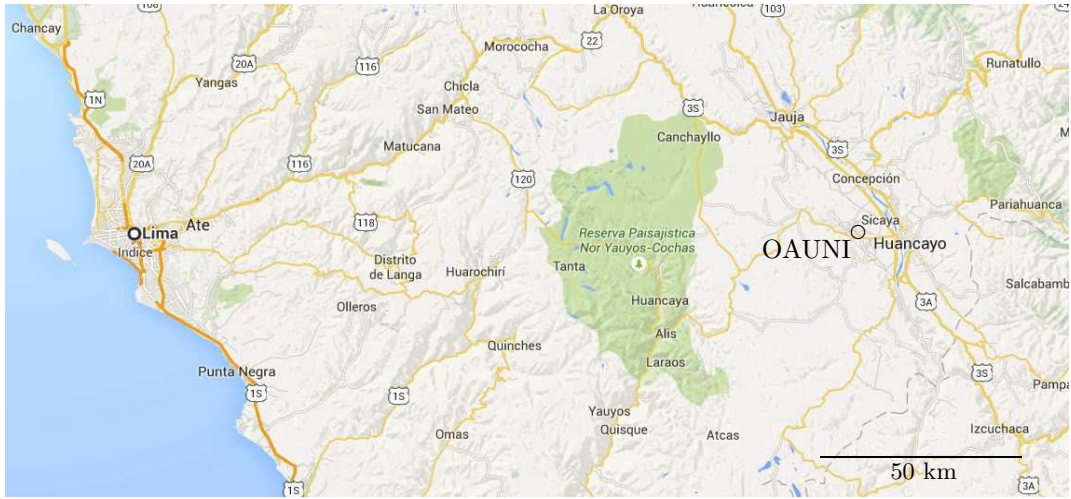}}
    \put(-136,100){\circle{5}}
    \put(-172,90){OAUNI}
    \put(-150,15){\line(1,0){75}}
    \put(-124,7){\small{50 km}}
\end{picture}
\captionof{figure}{\small{OAUNI (circle) is located to 315 km (by car) from Lima at the Peruvian Centarl Andes (3300 m.u.s.l ). OAUNI is installated inside the IGP's Huancayo Observatory (\textit{Google maps}).}}\label{mapa}
\end{figure*}

A collaboration between the Astronomy Area of the Geophysical Institute of Peru (IGP) and FC$-$UNI let the installation of OAUNI inside the Huancayo's IGP site at the peruvian central Andes (see Figure~\ref{mapa}). The logistic expenses for this installation were kindly covered by UNI General Research Institute (IGI) on the second semester of 2014.

In this work we describe in detail the current equipment available in OAUNI (Sect.~\ref{instru}), along with the recent installation phase at Huancayo (Sect.~\ref{instal}). The beginning of the commissioning phase and the first light also will be shown (Sect.~\ref{comis}). Finally, we comment the future perspectives of the project (Sect.~\ref{persp}).

\vspace{8mm}
\vspace{0.5cm}


\section{OAUNI instruments}\label{instru}

The main instrument currently available at OAUNI are the following (see Figure~\ref{rcos}):

\subsection{Optical tube}
The optical tube (OT) is the most important component of OAUNI. The OT is Cassegrain type with Ritchey-Chrétien design. The diameter of the primary mirror is 0.51~m and was manufactured by RC Optical Systems (RCOS). This optical design is optimized to eliminate the comma and yielding a wide field of view. Their hyperbolic primary and secondary mirrors help to this purpose. In addition, the system is free of spherical aberration and as a reflector system, the chromatic aberration is null. These characteristics make to this scope into the standard one at the professional observatories. The focal ratio f/8.2 of the system let to have a good compromise of wide field of view for survey-type applications.


\subsection{Robotic mount}

The pointing process of the scope is performed using an automatized mount. The OAUNI has a high precision Paramount ME II robotic mount. The payload capacity is 109~kg, being this enough to support the OT and the additional instruments (cameras and filters). Using a proper software (\textit{The Sky Pro}) which controls the mount, the pointing precision is no higher than 30". This let it to find astronomical objects in a simple and easy way.

\subsection{CCD camera}

The astronomical images are gathered digitally using a camera with a CCD detector. The main CCD camera of OAUNI is a STXL$-$6303E Santa Barbara Instruments Group (SBIG). This camera has a high sensibility CCD chip with a 9$\mu$m pixel size and an total area of 3072$\times$2048 pixels$^{2}$. This big size along with the OT focal ratio let to have a wide field of view of $\sim$23'$\times$15' with a plate scale of 0.45"/pixel.

The camera accepts a filter wheel (SBIG FW8G$-$STXL) which lets to use optical filters to separate the beam in particular wavelength intervals. The combination of the camera and its filter wheel offers the possibility to gather digital images up to  eight different filters. In addition, this wheel has a proper CCD detector, smaller than the main CCD, which lets auto-tracking during a given integration. For this procedure, a guide star is necessary on the collateral field of view. This, of course, turns to be more easy the tracking in long integrations.

\vspace{8mm}
\centerline{\includegraphics[width=8cm]{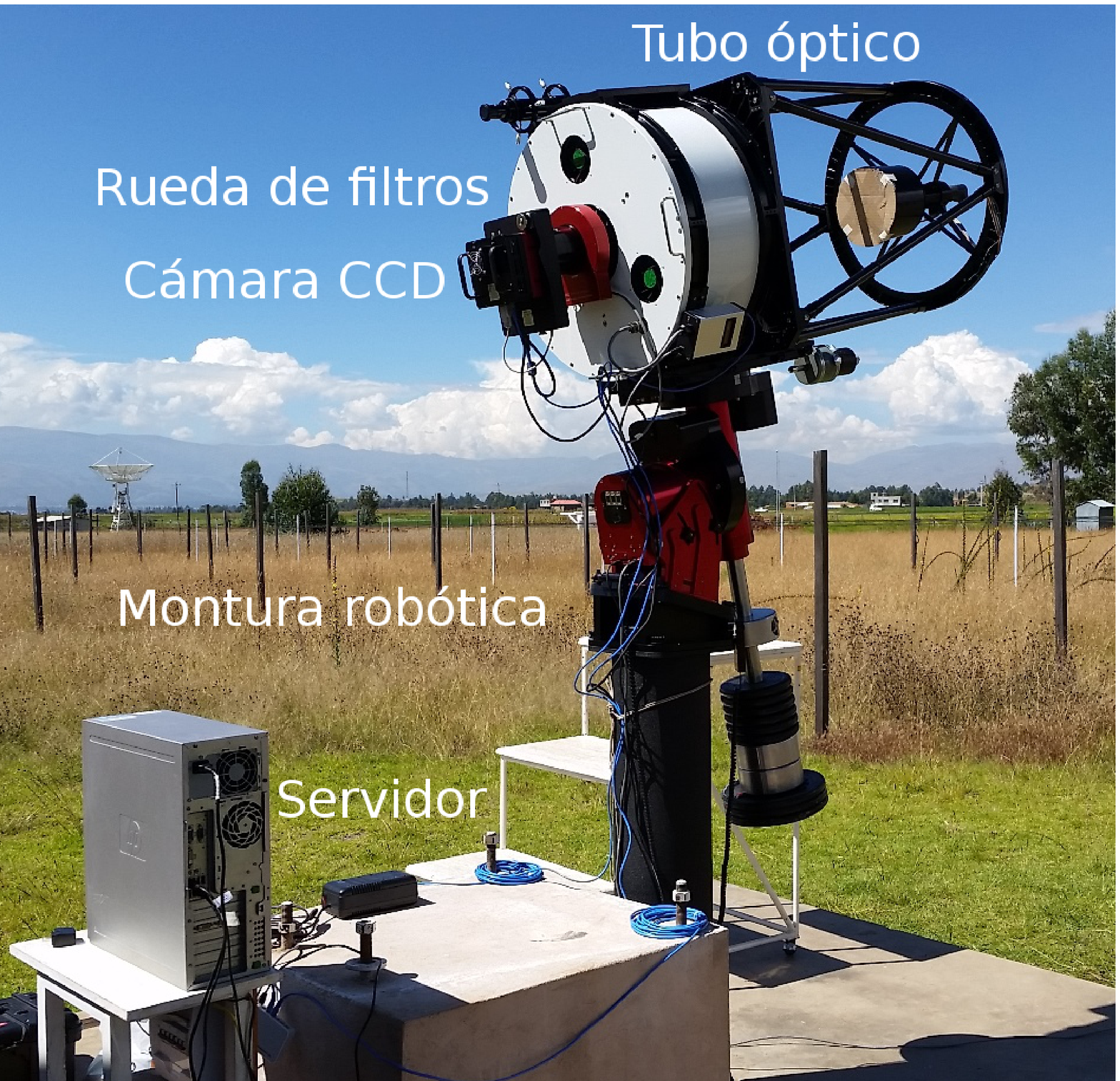}}
\captionof{figure}{\small{Main OAUNI equipments and instruments installed at the Huancayo Observatory.}}\label{rcos}
\vspace{0.5cm}

\subsection{Photometric filter set}\label{sistfil}

The OAUNI photometric filter system is the Johnson-Cousins (\textit{UBVRI}), which is the standard system for astronomical photometry (ver Figura~\ref{filters}). The five filters are properly inserted into the filter wheel letting to obtain stellar images in each broadband. The diameter of each of them is 50mm. Their quality is well tested to do research and the high transmission ($>$ 95\%) lets to observe dimm objects. The manufacturer of these filters is Astrodon Inc.

\vspace{8mm}
\centerline{\includegraphics[width=5cm]{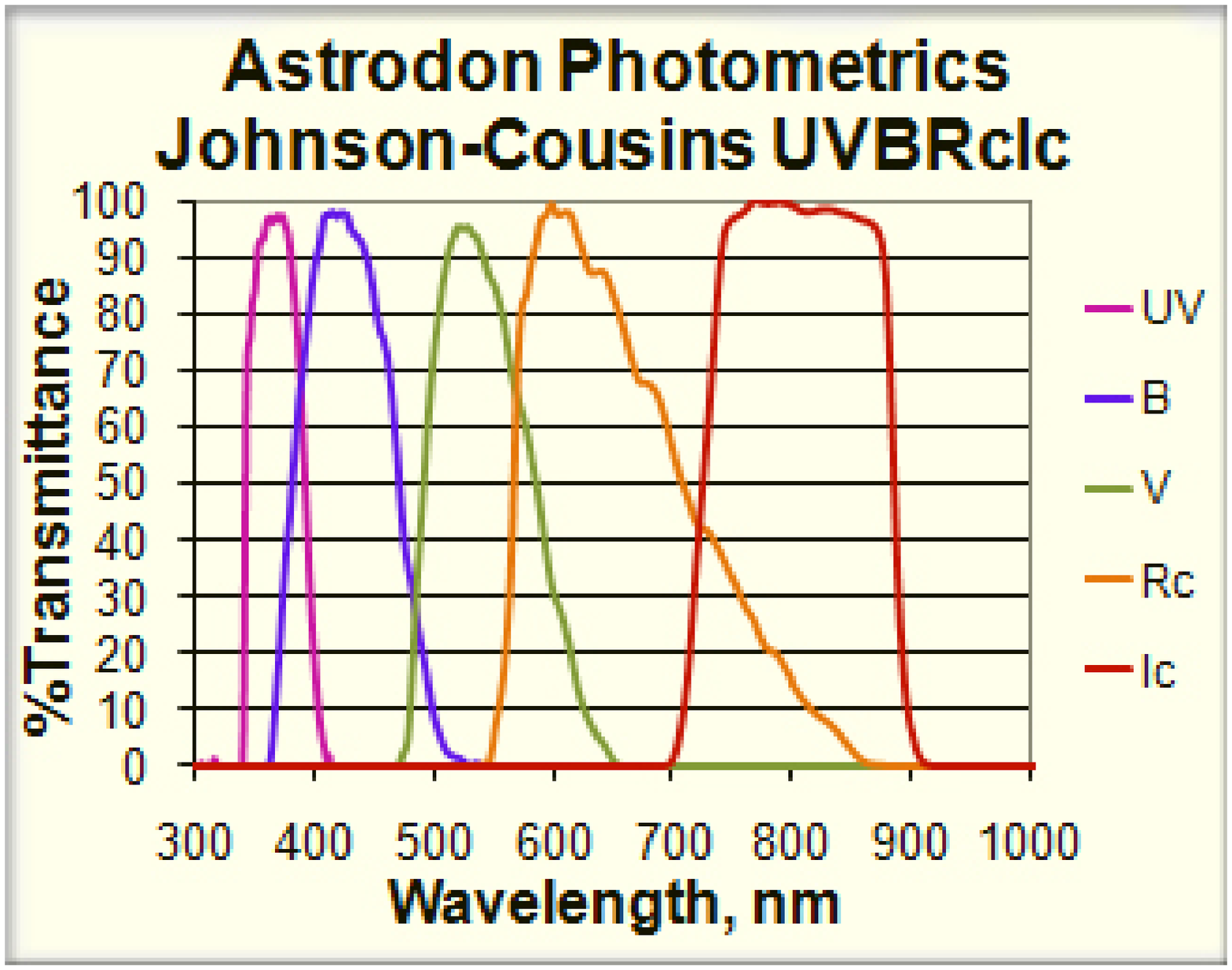}}
\captionof{figure}{\small{Transmittance curves of OAUNI photometric filter system (http://www.astrodon.com/uvbri.html).}}\label{filters}
\vspace{0.5cm}

\section{OAUNI Installation at Huancayo}\label{instal}

After the end of the pre-commissioning phase, we began the installation phase for the OAUNI equipments at the Huancayo Observatory (HO). This site belongs to the Geophysical Institute of Peru (IGP, in spanish). The transport was completed in several missions to HO during the last 2104 quarter and the first 2015 quarter.

\subsection{Mission 1}

In this first phase, the heavy duty steel pier was transported to HO. This pier is designed to support the mount and the optical tube. It was necessary to remove the old astrograph of its original foundation because the OAUNI was planned to be installed on this position. The size of the sliding roof of the enclosure that protected the old astrograph perfectly set the needed dimensions for the OAUNI installation. The foundation was conveniently prepared to support the new pier and special care was taken. In particular, the North-South direction was properly aligned with the new pier in order to facilitate the OAUNI sideral tracking. The alignment of the old astrograph helped to this purpose.

\subsection{Mission 2}

The next step of the installation process included the transport of the robotic mount and its latitude adjustment wedge. Special care was taken for the proper orientation of the mount. Its altitude angle was adjusted in order to align the mount axis with the  direction of South pole. The level precision device of the mount help to be more simple this procedure. In parallel, works were done to outfit the OAUNI enclosure of electricity supply.

\subsection{Mission 3}

In this mission, we performed several tests with the installed robotic mount using an eight inch telescope. It let us to get better the alignment of the mount axis with the North-South meridian. In this procedure was tested with success the pointing control software \textit{The Sky Pro}. In addition, the enclosure was equipped with proper security systems and it was started the conditioning of the control room (SECASI ambient), to 60m away. A local LAN network was satisfactorily tested to let the remote control of the scope and peripherals.

\subsection{Mission 4}

A crucial phase on the OAUNI installation was the transport of the 80 kg OT to Huancayo. A professional deliver was contracted to guarantee the success of this enterprise. After the OT arrived to HO, the careful OT engage on the robotic mount was planned. It was done using an hydraulic crane which let a slow but safe engage. New repairs to the telescope enclosure were done including several leaks on the old aluminium roof. The control room also was upgraded with new furniture.

\subsection{Mission 5}

Several additional peripherals were tested on the scope optical axis as the field rotator device. In addition, exhaustive tests were performed with the electronic control system of the scope. This system can control the focusing process on the secondary mirror, the temperature data record on both mirrors and the power supply for the primary mirror fans in order to thermalize the OT. New tests on transmission data using the LAN network were done between the telescope enclosure and the control room. Finally,  a new custom designed adapter was tested with sucess which let to engage the old ST7E CCD camera  to the OAUNI scope without problems.

\subsection{Mission 6}

Before this mission, the importation process of the astronomical filter set for the STXL$-$6303E camera (ver Sec.~\ref{sistfil}) was concluded. Then, all the 
 detection system (camara, filter wheel and filters) was transported to HO, where was properly installed. Local tests were performed indicating a perfect work of the mechanical and electronics components. The installation of a dedicated server (see Figure~\ref{rcos}) which controls the robotic mount also was done with success. Finally, with all the equipments and peripherals instalated on OAUNI, we proceed to the final weight balancing of the scope. Pointing and focusing tests were done with success and after that the installation process was concluded. In this point, the scope was ready to start the commissioning phase.

\vspace{8mm}
\begin{figure*}
\begin{picture}(0,300)
\centerline{\includegraphics[trim=0px 1px 1px 0px, clip=true,angle=0,width=16cm]{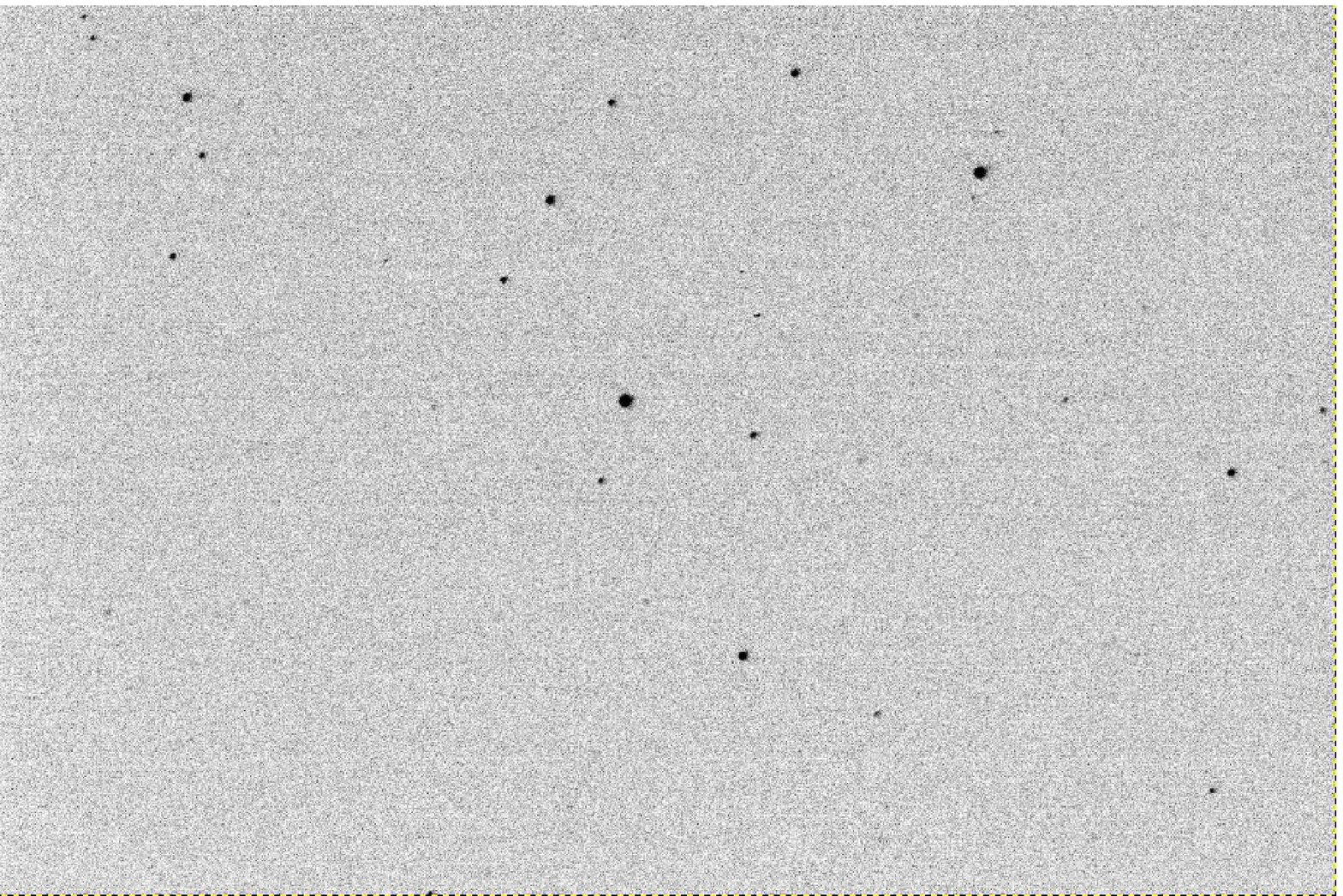}} \\
\put(-415,270){TYC6734-261-1 (11.30)}
\put(-263,166.5){HD123369 (9.50)}
\put(-144.5,244.5){HD123403 (9.81)}
\put(-208.5,278){TYC6734-353-1 (12.06)}
\put(-290,234){TYC6734-841-1 (11.40)}
\put(-223,79.5){TYC6734-363-1 (11.40)}
\put(-172,143){TYC6734-395-1 (11.99)}
\put(-472,90){\textbf{OAUNI}}
\put(-472,80){\textbf{RCOS 0.51 m Telescope}}
\put(-472,70){\textbf{Paramount MEII mount}}
\put(-472,60){\textbf{SBIG STXL6303E Camera}}
\put(-472,50){\textbf{Astrodon V filter}}
\put(-472,40){\textbf{Huancayo, Peru}}
\put(-472,30){\textbf{2015-05-20 20:05 (LT)}}
\put(-472,20){\textbf{Integration time: 5 s}}
\put(-50,20){\vector(0,1){30}}
\put(-50,20){\vector(-1,0){30}}
\put(-53.5,53){\textbf{E}}
\put(-89.5,16.5){\textbf{N}}
%
\end{picture}
\captionof{figure}{\small{OAUNI first light on 2015/05/20. The field stars are identified with their visual magnitude in parenthesis. The field of view is 23'$\times$15'.}}\label{primluz}
\end{figure*}
\vspace{0.5cm}

\section{Commissioning and First Light}\label{comis}

The first OAUNI observation of a stellar field was performed on 2015/may/20 at HO (see Figure~\ref{primluz}). This happened during the beginning of the OAUNI commissioning phase. The sky conditions for the first light were not the best ones for astronomical observations. Excessive relativity humidity ($\sim$70$\%$) and intermittent cloudy skies played against a better imaging quality. The measurement was gathered through the visual filter (\textit{V}) and also far from perfect focusing. However, a five seconds integration revealed well resolved stars up to 12 magnitude. The wide field of view yielded by the combination of OT and camera (RCOS 0.5m $+$ STXL6303E) show all its potential. In particular, the advantage to have a field rotator device let to orient the CCD detector to the equatorial system in a very easy way. The manual and by \textit{software} pointing systems were tested individually and both worked without problems.

The next commissioning phases will let to get better the pointing precision in order to get the nominal reference value ($<$ 30"). In addition, a better collimation of the OT will let to avoid unnecessary light losses on the focal point. New missions are programmed for the next semester (2015-II) given continuity to this process and also letting to start the science programs.

\section{Perspectives}\label{persp}

In the short term, the OAUNI project has as priority to develop 
photometric observational programs. In this sense, the precision of the photometric measurement is the extreme importance for the stellar variability programs. Below, we show an analysis of the first light image showing that the measurement precision has a good fit with the modeled values.

\vspace{8mm}
\centerline{\includegraphics[width=8cm]{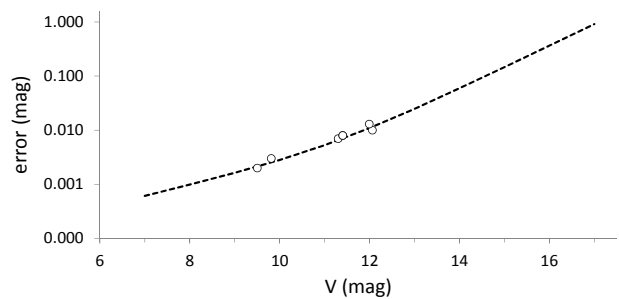}}
\captionof{figure}{\small{Simulation of the OAUNI photometric precision (error) considering the STXL6303E detector specifications (dotted line). It was assumed a integration time of 5 s, a seeing of 2" and a sky background of 20 mag/"$^{2}$. The photometric error computed directly from the first light image (Fig.~\ref{primluz}) also is shown (circles).}}\label{error}
\vspace{0.5cm}

Figure~\ref{error} shows a model for the photometric error including the specifications of our detector (gain = 1.5 e-/ADU, readout noise = \mbox{11 e- rms}). The assumed sky quality was just reasonable in order to be consistent with the observing conditions on the first light (\textit{seeing} = 2" and sky background = 20 mag/"$^{2}$). We can show that a photometric precision of 0.1 mag is possible for weak objects of 15 mag, using short integration times (IT = 5 s). This time was the same IT used on the first light. Obviously weaker objects could be observed with  this precision if proper and longer IT are used.

As we noted above, the first light image recorded well sampled objects up to 12 magnitude. For an astrometric identification of the field objects we used \textit{Astrometry.net}~\cite{lan10}. After that, the aperture photometry~\cite{wel94,dav94} performed over the identified objects (see Figure~\ref{primluz}) let us to compute the respective photometric errors. In this exercise we used the standard software for reduction of astronomical images with CCD (\textit{IRAF}\footnote{\textit{IRAF} is distributed by the National Optical Astronomy Observatory, which is operated by the Association of Universities for Research in Astronomy (AURA) under a cooperative agreement with the National Science Foundation.}). Figure~\ref{error} shows the photometric precision obtained from the analyzed data and clearly a good fit with the modeled values is found. This fact gives us trust that good quality photometric measurements can be performed with actual OAUNI setup.

The ongoing OAUNI commissioning phase also considers to make tests with an astronomical spectrograph. After it can be implemented, the option to perform scientific projects using low resolution astronomical spectroscopy will also be available at OAUNI. In addition, a fast CCD camera obtained in collaboration with The Paris Observatory also will be intensively tested in this phase. In a middle term, this camera will let to start the stellar occultations project using the OAUNI capabilities. In this sense, we estimate that the number of peripherals and equipments for the OAUNI project will be growing in the future. New financial sources are being explored for this purpose.

Finally, the observational astronomy courses at FC-UNI also will have support with OAUNI. In the middle term, we plan that last year students on Physics and/or Physics Engineering can complete initiation scientific activities at OAUNI. With this, undergraduate thesis will be defended in a routinely way with observational data gathered at OAUNI letting to grow the number of students involved in astronomy. Of course, this will encourage the next step in the academic live of these students including their postgraduate studies in this area. When this point will be reached, we consider that one of the main aims of the OAUNI project will be complete, namely, the astronomy development at UNI.

\section{\textit{Acknowledgments}}\label{agrad}

\small{The authors are grateful for the financial support of the UNI Rectorate, The World Academy of Sciences (TWAS), Faculty of Sciences - UNI, and UNI General Research Institute for the OAUNI project. We are also grateful to Dr. Pedro Canales for his firm support on the beginning of the project. A special gratitude to Drs. Abel Gutarra, Humberto Asmat, Armando Bernui, Susana Petrick and José Ishitsuka, along with Mg. Hugo Trigoso, for their support. A special mention of grateful to Mrs. Elena Ascanio, FC-UNI Secretary, for her invaluable support on the logistics of the project. Also, we want to thank to GA members: Erika Torre, Diego Berrocal, Mg. Guido Granda, and Dr. Nobar Baella for their support on the pre-commissioning phase. Finally, we are also grateful to the Huancayo Observatory personal for their help in the final installation phase of OAUNI.}

\end{multicols}

\begin{center}
-----------------------------------------------------------------------------------------------
\end{center}
\begin{multicols}{2} 
\small{

}
\end{multicols} 

\end{document}